\documentclass[twocolumn,aps]{revtex4}
\usepackage{graphicx}   
\usepackage{latexsym}   

\newcommand{\be}{\begin{equation}}
\newcommand{\ee}{\end{equation}}
\newcommand{\bea}{\begin{eqnarray}}
\newcommand{\eea}{\end{eqnarray}}

\newcommand{\lton}{\mathrel{\lower.9ex
                  \hbox{$\stackrel{\displaystyle <}{\sim}$}}}

\begin{document}
\title{$\phi$-Meson Production at RHIC, Strong Color 
Fields and Intrinsic Transverse Momenta}
\author{Sven~Soff$\,^{1,2}$, 
Srikumar Kesavan$^{2,3}$,
J\o rgen~Randrup$^2$, 
Horst~St\"ocker$^1$, 
Nu Xu$^2$}
\address{
$^1$Institut f\"ur Theoretische Physik, J.W.\ Goethe-Universit\"at,
60054 Frankfurt am Main, Germany\\
$^2$Lawrence Berkeley National Laboratory, 
Nuclear Science Division 70-319, 
1 Cyclotron Road, Berkeley, CA 94720, U.S.A.\\
$^3$Yale University, P.O. Box 204684, New Haven, CT 06520, U.S.A.}
\date{\today}
  
\begin{abstract}
We investigate the effects of strong color fields and of the associated 
enhanced intrinsic transverse momenta on the $\phi$-meson production 
in ultrarelativistic heavy ion collisions at RHIC.
The observed consequences include a change of the spectral slopes, varying 
particle ratios, and also modified mean transverse momenta. 
In particular, the composition of the production processes of $\phi$ mesons, 
that is, direct
production vs.\ coalescence-like production, depends strongly on 
the strength of the color fields and intrinsic transverse momenta and thus 
represents a sensitive probe for their measurement.
\end{abstract}
\pacs{PACS numbers: 25.75.-q, 25.75.Ld, 12.38.Mh, 24.10.Lx}
\narrowtext
\vspace*{-0.8cm}

\maketitle 

Ultrarelativistic heavy ion collisions provide a unique tool
to study elementary matter at energy densities so high that
a phase transition from partonic deconfined matter
to hadronic matter is predicted by QCD lattice 
calculations \cite{Karsch:2001vs}.
The properties of such a highly excited state depend
strongly on the initial conditions such as the specific entropy
density or the thermalization time.
A very interesting feature of relativistically colliding
heavy nuclei may be the production of strong color electric fields. 
Their existence implies a number of  characteristic consequences that 
are of utmost importance for the interpretation of experimental data. 
Among them are, for example, the predicted strangeness enhancement 
\cite{Rafelski:pu} and 
a strong baryon transport \cite{Mishustin:2001ib,Magas:2000jx,Soff:2002bn}. 

Here, we focus on the production of the
$\phi$ meson in Au+Au collisions at the 
BNL Relativistic Heavy Ion Collider ($\sqrt{s}_{NN}=200\,$GeV).  
The $\phi(s\bar{s})$ meson is expected to be a good messenger 
of the early stage of the evolution 
because its constituent quarks must have been
produced during the collision
(since they carry strangeness)
and this must have occurred early on (since they are massive)
\cite{Koch:1986ud,Shor:ui}.
Moreover, $\phi$'s have a rather 
small hadronic cross section with nonstrange hadrons and thus may leave the 
hadronic phase relatively unperturbed \cite{Shor:ui}. 
We will demonstrate that the $\phi$ meson is particularly sensitive 
to the color field properties. Their abundance depends strongly on the 
field strength. The transverse momentum generation by the strong color fields 
may lead to a substantial hardening of the spectra. The interplay between 
the mass and transverse momentum generation strongly influences 
the composition of the spectra. Directly produced $\phi$'s are shown 
to compete with a substantial contribution of $\phi$'s produced in 
coalescence-like $K\bar{K}$ collisions. The relative contributions 
are directly correlated to the magnitude of the color field strength and the 
associated intrinsic transverse momenta. 

Since the $\phi(s\bar{s})$ is analogous to the $J/\psi(c\bar{c})$
(they have the same quantum numbers: $I^G\,(J^{PC})=0^-\,(1^{--})$),
an improved understanding of in-medium $\phi$ production and decay 
may also elucidate the features of the $J/\psi$ dynamics.
For example, $J/\psi$ production by $D\bar{D}$ coalescence
is probably an important production mechanism at RHIC
\cite{Bratkovskaya:2003ux}
and corresponds to $K\bar{K}\to\phi$ coalescence.

The investigation is performed 
in the framework of a relativistic transport model 
(Ultrarelativistic Quantum Molecular Dynamics UrQMDv1.3) that is based on 
(di)quark, hadron (resonance), and string degrees of freedom \cite{urqmd}.
It simulates multiple interactions of ingoing and newly produced particles, 
the creation and fragmentation of color strings,
and the formation and decay of baryonic and mesonic resonances.
At RHIC energies, subhadronic degrees of freedom are
of major importance.
They are treated via 
the introduction of formation times for hadrons that are produced in the
fragmentation of strings \cite{Andersson:ia}.
Leading hadrons of the fragmenting strings contain the valence quarks
of the original excited hadron. 
They interact even during their formation time, with a reduced cross section
defined by the additive quark model,
thus accounting for the original valence quarks contained in that
hadron \cite{urqmd}.
Newly produced (di)quarks are only allowed to
interact after they have coalesced into hadrons.
Their formation times are inversely proportional to the
string tension $\kappa$ via $t_f \sim 1/ \kappa$.
The larger the string tension the shorter and short-living are the
strings with a certain total energy.
Secondary scatterings are important, for example, for transporting
baryon number from projectile and target rapidity closer to midrapidity.
The collision spectra are
largely dominated by (di)quark degrees of freedom \cite{Winckelmann:1996vu}.
Fully formed hadrons are involved
only in the very first collisions and at the later stage.  
(For further details of the model, see Ref.\ \cite{urqmd,Bleicher:2000pu}.)

Strong color electric fields, that is, an effectively
increased string tension in a densely populated colored environment of
highly excited matter leads to a modified particle production process 
in hadronic collisions. 
A background color electric field 
is formed between two receding hadrons which are color charged
by the exchange of soft gluons while colliding.
In nucleus-nucleus collisions the color charges may be considerably greater
than in nucleon-nucleon collisions due to the almost simultaneous
interaction of several participating nucleons \cite{kajantie85}.
This leads to the formation of strong color electric fields.
With increasing collision energy, the number and density of
strings grows so that they start overlapping, 
thus forming clusters that act as new effective sources 
for particle production \cite{biro84,sor92}.
The multiplicities
of, for example, strange baryons or antibaryons should be
strongly enhanced 
\cite{sor92,gyulassy90,gerland95,Soff:1999et,Soff:2001ae} once the color
field strength grows.
The abundances of (multiply) strange (anti)baryons
in central Pb+Pb collisions at CERN SPS \cite{and98a}, for example,
can be understood within
the framework of microscopic model calculations \cite{Soff:1999et,Soff:2001ae}
in terms of an enhancement of
the elementary production probability of $s\overline{s}$ pairs, 
which is governed in the string models \cite{Andersson:ia}
by the Schwinger mechanism \cite{schwing51}.
This corresponds either to
a dramatic enhancement of the string tension $\kappa$
(from the default $\sim 1\,$GeV/fm to $3\,$GeV/fm) or to
quark masses $m_q$ that are reduced from their constituent quark values
to current quark values as motivated by chiral symmetry
restoration \cite{Soff:1999et}.
A variation of the string tension from $\kappa=1\,$GeV/fm to $3\,$GeV/fm
increases the pair production probability of
strange quarks (compared to light quarks)
from $\gamma_s=P(s\overline{s})/P(q\overline{q})=0.37$ to 0.72.
Similarly, the diquark production probability is enhanced
from $\gamma_{qq}=P(qq\overline{q}\overline{q})/P(q\overline{q})=0.093$
to 0.45, leading to an effectively enhanced baryon-antibaryon pair production. 
In general, heavier flavors or diquarks ($Q$) are suppressed
according to the Schwinger formula \cite{schwing51} by
\begin{equation}
\gamma_Q=\frac{P(Q\overline{Q})}{P(q\overline{q})}= \exp\left(
-\frac{\pi (m_Q^2-m_q^2)}{ \kappa} \right)\,.
\end{equation}

The string tension can also be expressed through its relation
to the Regge slope $\alpha'$. Based on a rotational string picture
the string tension $\kappa$ follows from the Regge slope
$\alpha'$ as \cite{goddard,Wong:jf}
\begin{equation}
\kappa=\frac{1}{2 \pi  \alpha'}.
\label{alphakappa}
\end{equation}
The empirical value of the Regge slope for baryons is
$\alpha'\approx 1\,$GeV$^{-2}$ \cite{green87} 
which yields a string tension of approximately $1\,$GeV/fm.
However, the multi-gluon exchange processes dominated by Pomeron exchange
in high-energetic nucleus-nucleus collisions
are described by a Regge trajectoy
with a smaller slope of $\alpha_P'\approx 0.4\,$GeV$^{-2}$
\cite{veneziano,collins}.
According to Eq.\ (\ref{alphakappa}) this translates into a
considerably larger ({\it effective}) string tension 
$\kappa$ \cite{Soff:2002bn}.
The magnitude of a
typical field strength
at RHIC energies has been suggested to be as large
as $5-12\,$GeV/fm \cite{Magas:2000jx}
(as a result of collective effects related to quark-gluon-plasma formation).

The transverse momentum generation in the string models is 
adjusted to describe elementary reactions. 
The newly produced particles from the decay of color strings 
obey a Gaussian distribution 
$f(p_t) \sim \exp(-p_t^2/\sigma^2)$ with a width 
$\sigma=0.55\,$GeV/c$\,(\approx \sqrt(\kappa/\pi))$. 
The Schwinger mechanism, the quantum tunneling of quark-antiquark 
and gluon pairs in the background color field, is governed 
by energy (and not only rest mass). 
Hence, the probability of producing a particle with 
mass $m_i$ and transverse momentum $p_t$ in a color field of
strength $\epsilon$ follows from 
$P_i \sim \exp[-\pi(m_i^2+p_t^2)/\epsilon]$.
As a consequence, a stronger color field
also leads to larger transverse momenta. 
In the present study we explore a general factorization {\it ansatz},
\begin{equation}
P_i \sim \exp(-\pi m_i^2/\kappa) \cdot \exp(-p_t^2/\sigma^2)\ ,
\end{equation}
where the mass and transverse momentum scales, $\kappa$ and $\sigma$, 
are independent. 
This allows us also to disentangle and better understand the effects 
of strong color fields on particle production via the 
parameter $\kappa$ and on the transverse momentum generation 
via the intrinsic transverse momentum $\sigma$. 
An increased intrinsic transverse momentum has also been discussed, 
for example, in the context of the color glass condensate (CGC) model. 
The spectra of heavy ion data were shown to follow a simple 
scaling law in agreement with an  
intrinsic $p_t$ broadening \cite{McLerran:2001cv,Schaffner-Bielich:2001qj}. 
In the framework of the CGC picture 
the transverse momentum distributions of hadrons in heavy ion collisions
are basically determined through the initial momentum distributions of gluons
which, in turn, are controlled by the saturation of the initial
gluon density at high energies.
As a consequence, the CGC picture predicts that the
mean transverse momenta scale like the square root of the
charged particle multiplicity per unit rapidity and unit
transverse area ($\langle p_t \rangle \sim
\sqrt{ dN_{\rm ch}/(dy\, \pi R^2)}$).
Also for the description of direct photon production a substantial 
$p_t$ broadening effect is required \cite{Dumitru:2001jx}.
Similarly, an increased intrinsic transverse momentum can be deduced 
from dilepton spectra (in the Drell-Yan region) \cite{Gallmeister:2000ra}.  

\begin{figure}[htp]
\vspace*{-1cm}\hspace*{-0.4cm}
\includegraphics[angle=-0,width=9.5cm]{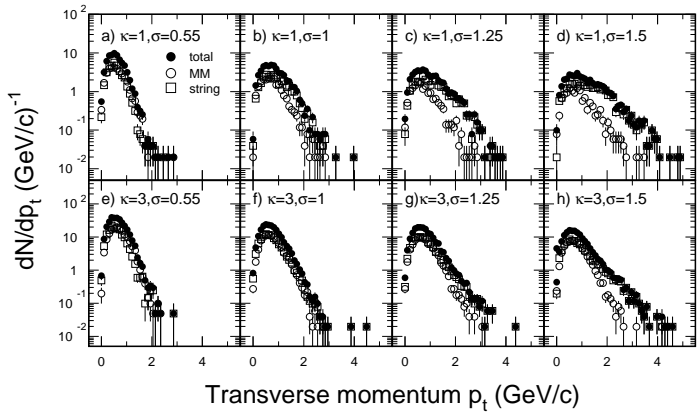}
\mbox{ }\vspace*{-1cm}
\caption{Transverse momentum spectra of $\phi$-mesons at midrapidity 
($|y|<1$) for central Au+Au collisions ($b<2\,$fm)  
at RHIC ($\sqrt{s}_{NN}=200\,$GeV). 
The different panels correspond to different values of the 
string tension $\kappa$ and the intrinsic transverse momentum $\sigma$. 
The spectra of all $\phi$'s (full circles) are compared 
to those which result from meson meson coalescence (open circles) 
and to $\phi$'s that are produced directly 
from string decays (open squares).}
\label{fig1}
\vspace*{-0.6cm}
\end{figure}

Figure \ref{fig1} 
shows the calculated transverse momentum distributions $dN/dp_t$ of 
$\phi$ mesons. The distributions are calculated for different values 
of the color field strength $\kappa$ (mass term) and the intrinsic transverse 
momentum $\sigma$. Moreover, the components of the spectra, 
i.e., $\phi$'s resulting from direct string decays and $\phi$'s 
produced via $K\bar K$ resonant scatterings are shown seperately. 
The relatively strong contributions from this coalescence-like 
production channel was already predicted for  
lower energies \cite{Soff:2001ae} and is also seen in other model 
calculations \cite{Pal:2002aw,Bravina:2002qg,Berenguer:1994qp}.   
For the vacuum parameterization, i.e., $\kappa=1\,{\rm GeV}/{\rm fm}$ and 
$\sigma=0.55\,$GeV/c, these two components contribute about equally
and the corresponding transverse momentum distributions 
exhibit the same slope. 
Increasing the intrinsic transverse momentum broadening parameter $\sigma$ 
leads to a stronger dominance of the direct production 
for the $\phi$-meson production. The slope of the total distribution 
is then largely dominated by the string channel, i.e., the 
coalescence mechanism contributes considerably less if 
the intrinsic transverse momentum is enhanced. 
In fact, the large values $\kappa$ and $\sigma$ describing the strong 
color fields lead to a power-law $p_t$  
distribution, most clearly seen in Fig.~\ref{fig1}(h). 
On average, kaons have larger (relative) momenta 
(in the case of increased intrinsic transverse 
momenta) and, hence, they are less likely to produce a $\phi$ meson.  

\begin{figure}[htp]
\vspace*{-1cm}   
\includegraphics[angle=-0,width=9.5cm]{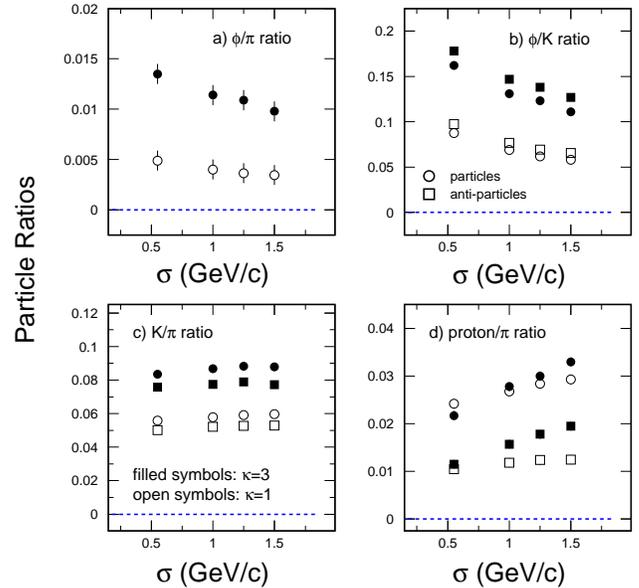}
\mbox{ }\vspace*{-1cm}
\caption{Particle ratios ($\phi/\pi,\, \phi/K,\, K/\pi,\,$and$\,p/\pi$ 
at midrapidity ($|y|<1$) as a function of the 
intrinsic transverse momentum $\sigma$ and for 
string tensions $\kappa=1\,$GeV/fm (open symbols) and $\kappa=3\,$GeV/fm 
(full symbols) for central Au+Au collisions ($b<2\,$fm)  
at RHIC ($\sqrt{s}_{NN}=200\,$GeV). Squares show the corresponding 
ratios for the antiparticles.}
\label{fig2}
\vspace*{-0.6cm}
\end{figure}

The particle ratios $\phi/\pi,\, \phi/K,\, K/\pi,\,$and$\,p/\pi$ are
shown in Fig.~\ref{fig2} as a function of the intrinsic transverse momentum
parameter $\sigma$ and for the vacuum color field strength
$\kappa=1\,$GeV/c and for strong color fields $\kappa=3\,$GeV/fm.
Clearly, the strange particle ratios are strongly enhanced
by increasing $\kappa$. The increased production probability
of $s\overline{s}$ pairs enhances the yields of (multiply) strange particles
relative to nonstrange particles.
Increasing the intrinsic transverse momentum $\sigma$ reduces
the $\phi/\pi$ and $\phi/K$ ratios.
This demonstrates again that $\phi$ production via $K\bar K$
coalescence becomes less important for larger
intrinsic transverse momenta.
The $p/\pi$ ratio follows an opposite trend. It increases
with $\sigma$. This is due to the hadronic final state interactions,
i.e., absorption processes which strongly depend on the collision energy.
The $p\overline{p}$ annihilation cross section increases strongly
toward small relative momenta.
In the $\kappa=3\,$GeV/fm case, the $p/\pi$ ratio increases also
faster than in the $\kappa=1\,$GeV/fm case.
More $p\overline{p}$ pairs are initially produced in the system
in the strong color field scenario but the corresponding
increased phase space densities of protons and antiprotons
and shorter formation times
(more scatterings) also lead to more (re)absorption processes.

\begin{figure}[htp]
\vspace*{-1cm}\hspace*{-0.4cm}
\includegraphics[angle=-0,width=9.5cm]{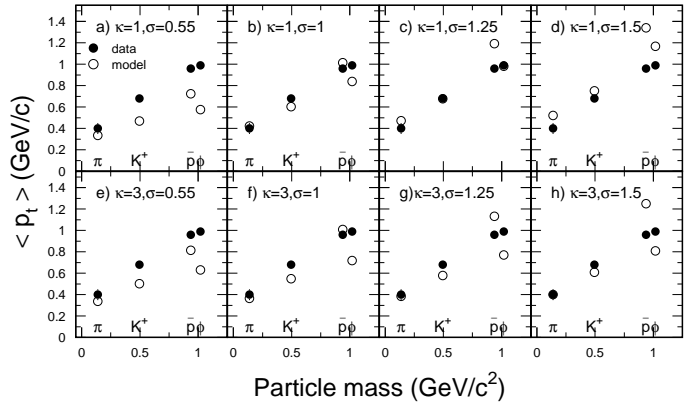}
\mbox{ }\vspace*{-1cm}
\caption{Mean transverse momentum $\langle p_t \rangle$
as a function of particle mass at midrapidity in central
Au+Au collisions ($b<2\,$fm)
at RHIC ($\sqrt{s}_{NN}=200\,$GeV).
The different panels correspond to different values of the
string tension $\kappa$ and the intrinsic transverse momentum $\sigma$.
The calculations (open circles) are compared to experimental data
(full circles).}
\label{fig3}
\vspace*{-0.6cm}
\end{figure}

Figure~\ref{fig3} 
shows the calculated mean transverse momenta of pions, kaons, protons, and
$\phi$ mesons in comparison to experimental data \cite{Adler:2002xv}. 
The mean transverse momenta naturally grow with the intrinsic transverse
momentum $\sigma$. The vacuum parameters $\kappa=1\,$GeV/fm and
$\sigma=0.55\,$GeV/c underpredict the mean transverse momenta.
Better agreement is obtained with an increased broadening parameter $\sigma$.
While the experimental values for protons and $\phi$'s are pretty close
to each other the calculated values for $\phi$'s are smaller than
those of protons. This can be attributed to the fact that the experimental
$\phi$ yields shown here are determined by the reconstruction via the invariant
mass of $K^+K^-$ pairs. Since the decay products of $\phi$'s, i.e., the
kaons, can suffer subsequent scatterings and thus get lost to
this experimental identification criterion the apparent yields are
considerably reduced. These rescatterings are more likely to happen 
at low momenta. As a consequence the experimentally observed (via $K^+K^-$)
mean transverse momenta of $\phi$'s are larger than
the ones of all $\phi$'s (accessible through the dielectron decay channel) 
including those whose decay products rescatter.
The hadronic and leptonic decay channels 
thus neccessarily lead to {\em different} spectral slopes and yields. 
This appears to be supported by preliminary data from the PHENIX 
\cite{Mukhopadhyay:2002bj} and STAR \cite{Adler:2002xv} collaborations. 
Previously, data at the SPS ($E_{\rm lab}=160\,A$GeV) from 
the NA49 \cite{Afanasev:2000uu} and NA50 \cite{Abreu:jq} collaborations 
showed different results in the hadronic and dimuon decay channels, 
respectively. 
These differences, however, could only partially be attributed to 
the rescattering of the decay products \cite{Soff:2001ae,Johnson:2001fv}. 
Moreover, calculations of a hadron gas at temperatures close to the 
phase boundary also yielded substantial rescattering for the $\phi$ mesons 
themselves \cite{Alvarez-Ruso:2002ib}. 
The comparison of the two decay channels may help to probe the strength of 
the color fields because the 
overall spectrum and the 
losses in the hadronic channel are
sensitive to the explicit values 
of the color field and to the intrinsic transverse momenta.  
In addition, the initial $\phi$'s (those that are possibly 
created through strong color fields) will be visible primarily in the 
dimuon channel \cite{Pal:2002aw}.

In summary we have shown that the combined effects of 
strong color fields as the enhanced production of heavier masses 
as well as increased intrinsic transverse momenta have 
a large impact on the hadronic observables, in particular for $\phi$ mesons. 
Their abundance depends strongly on the
field strength. The spectra experience a substantial hardening 
due to the presence of strong color fields. Moreover, we have 
demonstrated that the interplay between
the mass and transverse momentum generation strongly influences
the composition of the spectra. Directly produced $\phi$'s have been shown
to compete with a substantial contribution of $\phi$'s produced in
coalescence-like $K\bar K$ collisions. The relative contributions
are directly related to the magnitude of the color field strength 
and the associated intrinsic transverse momenta.
The measurement of both the hadronic and the leptonic decay channels 
of $\phi$ mesons, and the analysis of the expected differences between them, 
should help to substantially improve
our understanding of the production processes and the underlying dynamics 
which are both strongly influenced 
by the properties of the strong color fields.
\vspace*{-0.8cm}
\acknowledgements 
\vspace*{-0.4cm}
We are grateful to V.\ Koch and E.\ Yamamoto 
for valuable comments. We thank the
UrQMD collaboration for permission to use the UrQMD transport model.
S.S.\ has been supported in part by the Humboldt Foundation.
This research used resources of the National Energy Research Scientific 
Computing Center. 
This work is supported by the U.S. Department of Energy under Contract 
No. DE-AC03-76SF00098, the BMBF, GSI, and DFG.
\vspace*{-0.6cm}
\textheight 25.3cm

\end{document}